# Performance Analysis of DiffServ based Quality of Service in a Multimedia Wired Network and VPN effect using OPNET

Muhammad Aamir[1], Mustafa Zaidi[2] and Husnain Mansoor[3]

[1] MS-IT (Computing), SZABIST
Karachi, Pakistan
*aamir.nbpit@yahoo.com*

[2] Department of Computer Science, SZABIST
Karachi, Pakistan
*mustafainisb@gmail.com*

[3] Department of Computer Science, SZABIST
Karachi, Pakistan
*husnain.mansoor@szabist.edu.pk*

**Abstract**

Quality of Service (QoS) techniques are applied in IP networks to utilize available network resources in the most efficient manner to minimize delays and delay variations (jitters) in network traffic having multiple type of services. Multimedia services may include voice, video and database. Researchers have done considerable work on queuing disciplines to analyze and improve QoS performance in wired and wireless IP networks. This paper highlights QoS analysis in a wired IP network with more realistic enterprise modeling and presents simulation results of a few statistics not presented and discussed before. Four different applications are used i.e. FTP, Database, Voice over IP (VoIP) and Video Conferencing (VC). Two major queuing disciplines are evaluated i.e. 'Priority Queuing' and 'Weighted Fair Queuing' for packet identification under Differentiated Services Code Point (DSCP). The simulation results show that WFQ has an edge over PQ in terms of queuing delays and jitters experienced by low priority services. For high priority traffic, dependency of 'Traffic Drop', 'Buffer Usage' and 'Packet Delay Variation' on selected buffer sizes is simulated and discussed to evaluate QoS deeper. In the end, it is also analyzed how network's database service with applied Quality of Service may be affected in terms of throughput (average rate of data received) for internal network users when the server is also accessed by external user(s) through Virtual Private Network (VPN).

## 1. Introduction

Networks are now dealing with high-bandwidth traffic and applications having strict requirements of successful packet delivery with minimal delay and delay variations. Major applications may include Voice over IP (VoIP) and Video Conferencing (VC) which are highly sensitive to loss, delay and jitter [1].

When high-bandwidth and delay sensitive services are the part of network, some Quality of Service (QoS) mechanism is applied to guarantee successful packet delivery with reduced latency and jitter according to assigned priority of packets. A queuing discipline is implemented according to which packets are processed. Major queuing disciplines are First In First Out (FIFO), Priority Queuing (PQ) and Weighted Fair Queuing (WFQ) [2].

In FIFO, a single queue is maintained for all traffic and first packet arriving at the router is processed first. In PQ, individual queues are maintained for each priority level and router processes the complete queue with the highest priority first. Then, second highest priority queue is processed and so on. However, during the processing of packets in a certain queue, when packet(s) arrive at the higher priority queue, current processing is postponed and higher priority queue is again treated first. In this way, higher priority packets are always processed on priority in order to maintain successful packet delivery and uninterrupted services of delay sensitive applications. In WFQ, weights are assigned to each queue according to priority level. In this way, even lower priority packets can also get their share in the link's bandwidth without having to undergo unacceptable delays in processing due to higher priority packets [1, 2]. Weighted Fair Queuing is widely supported in current QoS implementations and configured with developed scheduling mechanisms such as Deficit Weighted Round-Robin (DWRR) and Negative-Deficit Weighted Round-Robin (N-DWRR) [3, 4].

"DiffServ" model is explained by Differentiated Services Code Point (DSCP) which is used to assign priority to a packet within IP header according to the type of service. It is widely used in today's QoS enabled IP communication systems and supersedes earlier specifications of marking of packets which used Type of Service (ToS). In addition, DiffServ has also an edge over earlier "IntServ" model of QoS which works through Resource Reservation Protocol (RSVP) by reserving network resources [5, 6].

DiffServ is widely used in current Internet-based communications where Quality of Service is ensured in such a way that border nodes mark packets according to their classification and possibly condition them. It is done through three most significant Class Selector bits. The packets are further treated by core routers where assigned DSCP value is used for each packet to forward it according to per hop behavior (PHB) associated with given DSCP value. Three bits next to class selector bits are used to manage drop precedence of the respective class. In this way, six bits in IP header are used to accomplish DiffServ based Quality of Service. The two least significant bits in 'Type of Service' byte are for the use of Explicit Congestion Notification (ECN) in routers which are capable to handle Active Queue Management (AQM). As a matter of fact, QoS is in high demand in Internet-based communications; however end-to-end QoS is still a big issue due to the default Best Effort paradigm of Internet architecture [3, 7, 8, 9, 10, 11].

The Virtual Private Network (VPN) [12] is a widely used technology through which users can access an organization's private network through public network (such as Internet) with assigned privileges. When services are accessed over Internet, they usually experience additional latency and delay due to firewall devices where proxy may induce additional processing delays [13]. VPN has in-built security features which make it a secure way to remotely access specific servers [14]. However, just as an increase in internal users affects the data rate (throughput) of a service experienced by other users, assigning access to external user(s) also produces an impact on the data rate of a service which may be experienced by internal users accessing the same server [15].

Researchers have done considerable work on Quality of Service both for wired and wireless IP networks. Recent research is focused to evaluate and improve QoS performance over IPv6 networks [16] and more consideration is given to high-bandwidth and delay sensitive services like Voice and Video [17]. Valuable work is being done towards the enhancement of end-to-end QoS in Internet-based multimedia communications [11]. Analysis on queuing disciplines may be found in [1], [2], [6] & [16] for the assessment of QoS performance in given networks under specific configurations.

## 2. Simulation Tool & Statistics

OPNET is used as the simulation tool in this study for required analysis. The tool provides several benefits to researchers and learners in terms of flexible topology design, availability of numerous applications, the validity of results and realistic analysis of performance measures in networks [18]. In this simulation study, statistics of queuing delay, queue delay variation (jitter), end-to-end delay & jitter, traffic drop, buffer usage and average data rate (throughput) are collected. These metrics describe the strength of QoS in a network [16].

## 3. Network Model & Configurations

In order to analyze queuing disciplines, a hypothetical network topology is considered in this simulation as shown in figure 1.

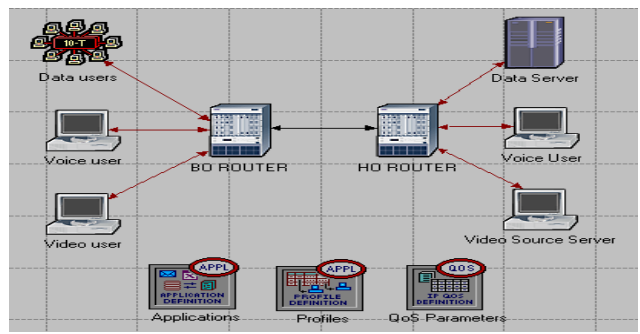

Fig. 1  Network Topology for Simulation.

In figure 1, HO Router is "Head Office Router" and BO Router is "Branch Office Router." Queuing disciplines are analyzed at HO Router's interface managing outbound traffic for BO Router. 'Data users' is a node of Local Area Network (LAN) having 10 users accessing FTP & Database applications over a 10 Base-T link. Therefore, we have a more realistic model of network as compared to earlier work [2]. The other users are accessing Voice over IP (VoIP) and Video Conferencing (VC) applications as mentioned.

For simulation discussed in this paper, G.711 encoding is used for voice application. For video, frame size of 128x120 pixels is configured with frame inter-arrival time = 10 frames/second. Database application works on Transaction Size of 200 bytes (constant) and Transaction Inter-arrival time is set to 30 seconds (exponential). FTP application works on File Size of 1000 bytes (exponential) and Inter-Request time is set to 3600 seconds (exponential).

Differentiated Services Code Point (DSCP) values are assigned to packets in OPNET as follows:

Table 1: DSCP values assigned to applications

| Application | DSCP Name | DSCP Value |
|---|---|---|
| Voice over IP (VoIP) | EF | 101110 |
| Video Conferencing (VC) | AF41 | 100010 |
| Database | AF21 | 010010 |
| FTP | DF (CS0) | 000000 |

In table 1, applications are mentioned from top to bottom with decreasing priority. EF is "Expedited Forwarding" which corresponds to telephony service. It has hard guarantee on delay and its variants [6]. AF41 is "Assured Forwarding" for multimedia conferencing service. AF21 is for client/server transactions and CS0 corresponds to standard class where the service is undifferentiated [19].

## 4. Simulation Results & Analysis

4.1 Queuing Delays

'Queuing Delay' is the amount of delay for which a packet has to wait in respective queue before it gets processed. Together with propagation delay and processing delay, it forms total delay or end-to-end delay for a packet [16]. In this simulation study, queuing delays experienced by all services are observed under both PQ & WFQ disciplines.

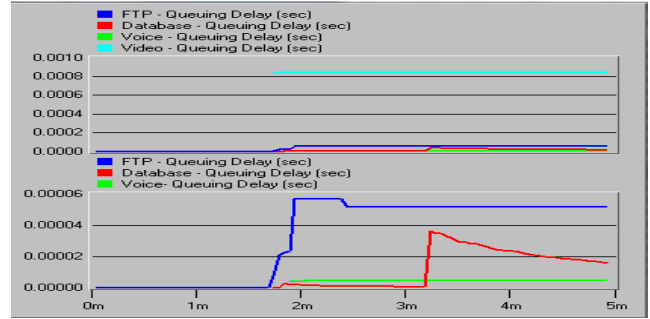

Fig. 2 Queuing Delays under PQ.

Figure 2 depicts that Video Conferencing service has the highest queuing delay under Priority Queuing (PQ) discipline for given configurations. Although video is at second highest priority in the network, high amount of packet payload has brought its delay at the highest level which is considerably more than other services. It shows that more time is consumed in servicing video packets which causes high delays. It has a definite impact on other services as well, particularly lower priority traffic. On the other hand, Voice over IP service has the lowest queuing delay as desired.

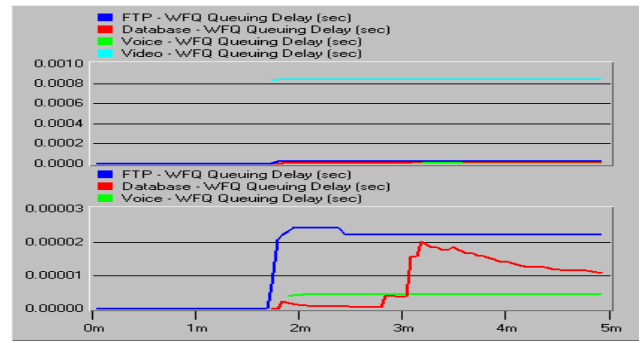

Fig. 3 Queuing Delays under WFQ.

Figure 3 shows that Video Conferencing service has the highest queuing delay under Weighted Fair Queuing (WFQ) discipline for given configurations. Voice over IP service has the lowest queuing delay.

4.2 Jitters

'Queue Delay Variation' or 'Jitter' is the variation in delay experienced by packets [16]. The existence of high jitter can lead a service to failure in a network, particularly the service which is highly sensitive towards delay. In this simulation study, jitters experienced by all services under both PQ & WFQ disciplines are observed.

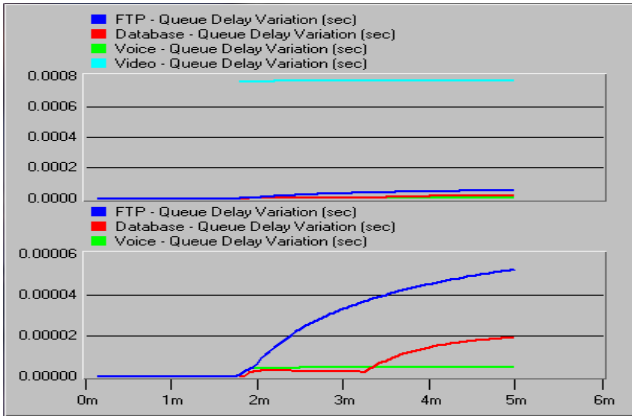

Fig. 4 Jitters under PQ.

Figure 4 shows that Video Conferencing service has the highest queue delay variation or jitter under Priority Queuing (PQ) discipline for given configurations, which is considerably higher than other services. Voice over IP service has the lowest jitter.

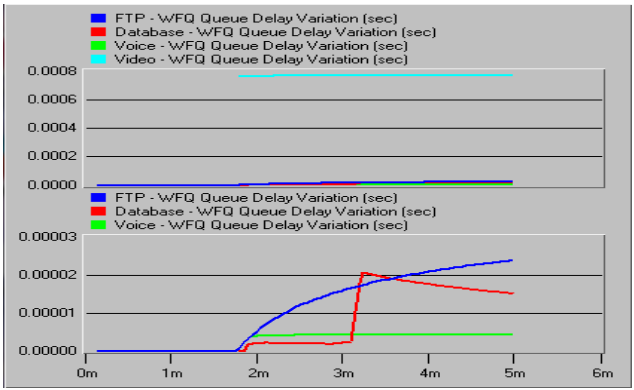

Fig. 5 Jitters under WFQ.

Figure 5 shows that Video Conferencing service has the highest queue delay variation or jitter under Weighted Fair Queuing (WFQ) discipline for given configurations. Voice over IP service has the lowest jitter.

4.3 Analysis of FTP Service

Comparison is done for the performance of individual services in the network for delay & jitter and it is examined that FTP service, under given configurations, experiences higher queuing delay and jitter in PQ discipline than WFQ. Therefore, WFQ is a better queuing discipline for FTP traffic in the network.

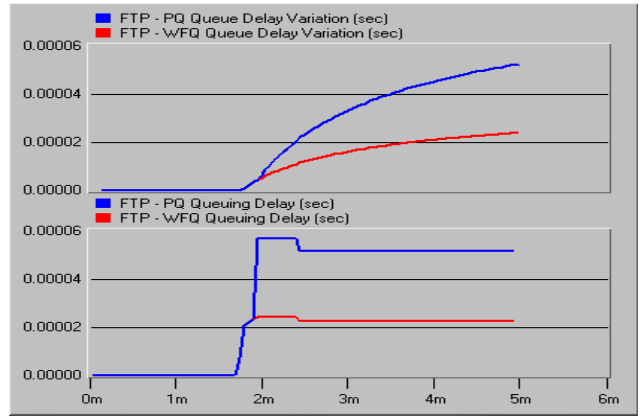

Fig. 6 Queuing Delay and Jitter under PQ & WFQ in FTP Traffic.

Figure 6 depicts that FTP service experiences more delay and jitter in PQ discipline because higher priority traffic is serviced first for entire queues. On the other hand, in WFQ discipline, even low priority traffic like FTP has its weighted share over available bandwidth which significantly improves its performance in terms of delay and jitter.

4.4 Analysis of Database Service

It is examined that Database service, under given configurations, also experiences higher queuing delay and jitter in PQ discipline than WFQ. Therefore, WFQ is a better queuing discipline for Database traffic in the network.

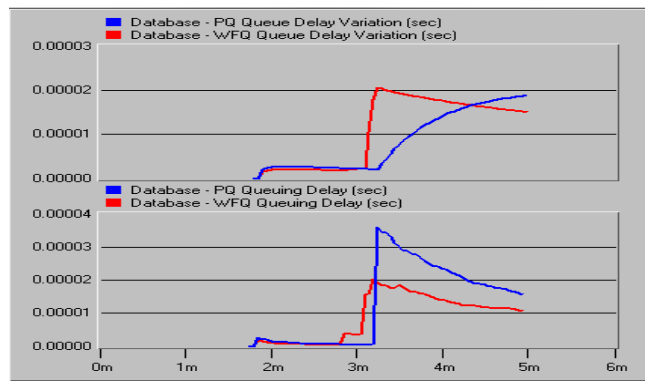

Fig. 7 Queuing Delay and Jitter under PQ & WFQ in Database Traffic.

Figure 7 shows that Database service also experiences more delay and jitter in PQ discipline like FTP because higher priority traffic is serviced first for entire queues. On the other hand, in WFQ discipline, Database service has improved performance in terms of delay and jitter. It has higher queue delay variation or jitter in WFQ than PQ initially, but with the passage of simulation time, it shows experiencing lesser jitter in WFQ whereas jitter in PQ increases.

4.5 Analysis of VoIP & VC Services

It is also analyzed that both VoIP and VC services, under given configurations, experience same queuing delays and jitters in respective PQ & WFQ disciplines (as evident from figure 2 to figure 5). This behavior of voice and video is obtained since they are high priority traffic in the network and thus bear hard QoS guarantees for delay and its variants in both PQ and WFQ disciplines. Earlier work in this direction [2] points out that WFQ is better for streaming video's queuing delay and PQ is better for its jitter. However, our tested scenario is configured with such values of traffic generation that it produces same delay and jitter in both PQ and WFQ. Therefore, it indicates that same delay and jitter may be observed for high priority traffic in both PQ and WFQ disciplines when clearly distinct service classes are present in a network. Voice has the lowest delay and jitter amongst all other type of traffic in the network. The voice packet is not very long but it is highly sensitive to delay and jitter due to the nature of service. The priority of voice in QoS configuration is therefore the highest. If its priority is brought down, the end user will receive low grade and broken voice [20]. On the other hand, video has the highest delay and jitter in the network being a service with high payload packets. Therefore, live streaming video transmission demands high quality link as well as disturbance less connection [2]. In this analysis, it is observed that queuing delay of video is the highest amongst all network traffic which is about 0.85 milliseconds. According to ITU-T specifications, delay-sensitive services should have delay of 150ms or less for unaffected service quality. There is no specific limit for jitters [8, 16].

4.6 Video – Traffic Drop & Buffer Usage

It has been analyzed that WFQ [2] is a better discipline for network's low priority services under configured values and defined priority settings. This discipline is further examined with variations for our high priority traffic i.e. voice and video. Therefore, further analysis is performed by introducing 'Traffic Drop' in the network through decreasing buffer size of router's interface and examining the 'Buffer Usage' statistic for video packets.

'Buffer' is memory location within router where packets are placed in queues before they get processed upon their turn [21]. By varying buffer size, we actually increase or decrease the queue length of a queuing discipline. High buffer size corresponds to a longer queue. When queues are completely filled and more packets arrive (case of buffer overflow), newly arrived packets are dropped by the router. Increased buffer usage indicates that more packets are being placed in queue(s) for processing. It strictly depends on the availability of space in buffer to accommodate the arriving packets.

Several mechanisms have been proposed and implemented to manage the queue length upon congestion in the network. Drop Tail (DT) mechanism was earlier used in which packets are dropped from the tail(s) of full queue buffer(s). Random Early Detection (RED) is mostly used at current times with improved implementations [22]. It is an Active Queue Management (AQM) algorithm which is based on the idea that incoming packets can be dropped by the router even if there is some space available in buffer(s) and router can send congestion signals to Transmission Control Protocol (TCP) before actual congestion occurs. In this way, queues and delays are prevented to grow too high [23]. In this simulation study, RED algorithm is enabled for the observed queuing disciplines.

In this analysis, five different buffer sizes are selected to examine different levels of packet drop and corresponding buffer usage for video conferencing service.

Table 2: Selected buffer sizes for video packets

| *Application* | *Buffer Size (Kilobytes)* | *Traffic Drop observed (Yes/No)* |
|---|---|---|
| Video Conferencing | 1 | Yes |
| Video Conferencing | 3 | Yes |
| Video Conferencing | 5 | Yes |
| Video Conferencing | 9 | Yes |
| Video Conferencing | 10 | No |

In table 2, buffer size of 10 KB indicates that there is no drop in video traffic at this value. It is selected for analyzing 'Buffer Usage' when there is no drop.

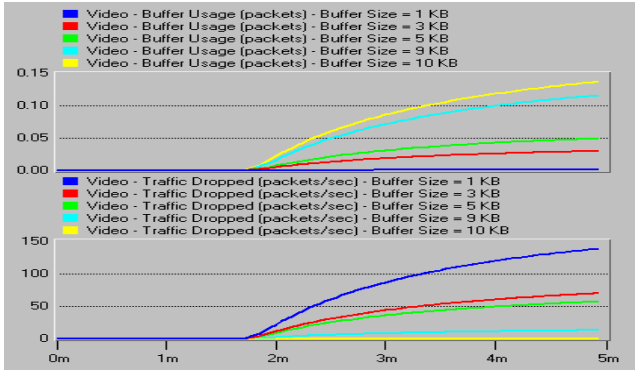

Fig. 8 Traffic Drop and Buffer Usage in Video Traffic.

In figure 8, it is observed that with increasing Buffer Size, as 'Buffer Usage' increases, corresponding 'Traffic Drop' is decreased. It happens because queue length is increased with higher value of buffer size and router is capable to place more packets in the buffer for processing. For given values, maximum buffer usage is observed at 10 KB buffer size when corresponding drop is zero.

4.7 Video – Buffer Usage & Packet Delay Variation

Analysis of 'Packet Delay Variation' statistic of video traffic is performed in the network and comparison is made with corresponding buffer usage.

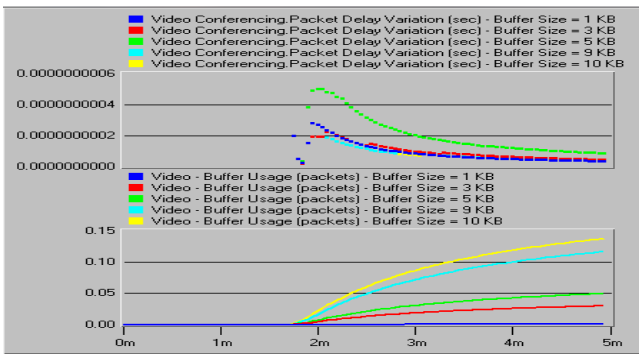

Fig. 9 Buffer Usage & Packet Delay Variation in Video Traffic.

In figure 9, we have end-to-end packet delay variation or jitter for Video Conferencing service in the network. It is presented in conjunction with corresponding buffer usage at different buffer sizes.

It is observed that with increasing buffer usage, end-to-end jitter in video is also increased when there is significant traffic drop. However, when we arrive at buffer sizes of 9 and 10 KB where traffic drop is very less and zero respectively, significant improvement in packet delay variation is seen. It is on minimum at 10 KB buffer size and found maximum at 5 KB.

4.8 Voice – Buffer Usage & Packet Delay Variation

Analysis of 'Packet Delay Variation' of voice traffic is also performed in the network and comparison is made with corresponding buffer usage.

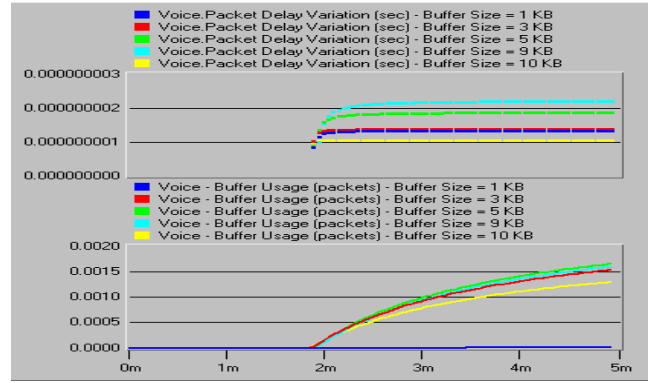

Fig. 10 Buffer Usage & Packet Delay Variation in Voice Traffic.

For voice, in figure 10, it is observed that with increasing buffer usage, end-to-end packet delay variation or jitter is also increased. However, at 10 KB buffer size, the buffer usage is medium. Voice has the highest priority in the network with Expedited Forwarding configuration [6]; therefore traffic drop is zero at all buffer sizes. When we arrive at buffer size of 10 KB where buffer usage is medium, significant improvement in packet delay variation is seen where it becomes at its minimum.

4.9 Database Traffic with & without VPN

Database is one of the services in simulated network. Usually at an enterprise level, database access is also provided to external users for certain query processing or similar work. The external users may be employees or customers. A secure way to give such an access to external users is Virtual Private Network (VPN). In this analysis, the network model with applied Quality of Service is extended to include Internet-based communication. In the extended network model, we include internet cloud and other necessary network devices i.e. Firewall, Routers and remote user stations. VPN configuration is applied to provide remote user access of 'Data Server' for accessing Database service.

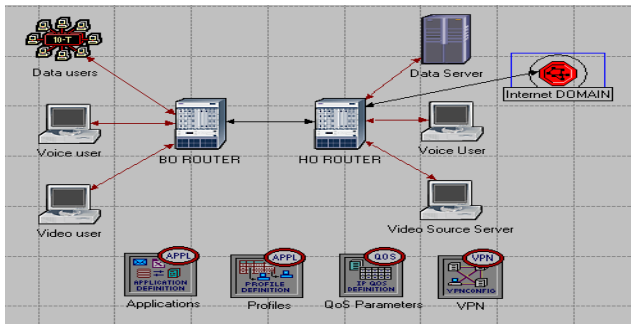

Fig. 11 Extended Network Model.

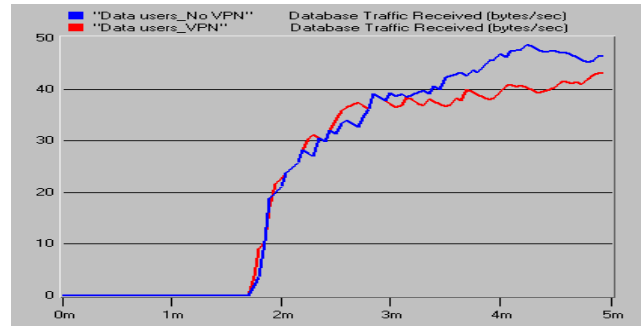

Fig. 13 VPN effect on Database Traffic for Internal Users.

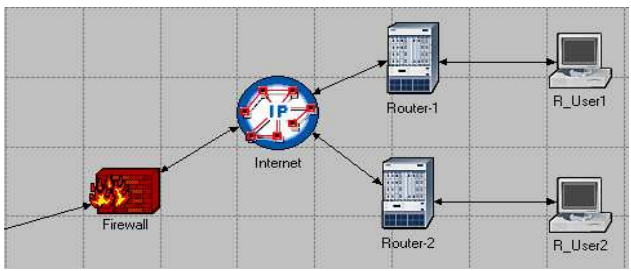

Fig. 12 Components of Internet DOMAIN.

In figure 11, the network model is extended by including 'Internet DOMAIN' and VPN configuration. In figure 12, components of Internet DOMAIN are presented. VPN is configured in such a way that Remote User 'R_User1' is allowed to access 'Data Server' for Database service whereas the same service is blocked for 'R_User2'. Therefore, the simulation result is obtained by adding the access of one remote user in the network. The complexity of network increases when QoS is applied on Internet traffic. The related work in this direction usually highlights QoS guarantees required by specified services. It is not evaluated in this paper whether QoS should be available on VPN based traffic or not. A simple VPN tunnel is established without QoS.

'Throughput' [21] is the average rate of messages successfully delivered over the network in unit time. The 'message' in network traffic scenario may be considered as a byte or packet. In this simulation, 'Database traffic received' by 'Data users' is analyzed before and after VPN implementation with mentioned setup of remote access.

In figure 13, it is observed that average data rate or throughput (in bytes/second) is decreased for internal network users i.e. 'Data users' after outside user is allowed to access network's 'Data Server' for Database service over Virtual Private Network. Therefore, it can be said that even QoS enabled services may experience throughput issues for internal users when servers are allowed to be accessed by external user(s) without specific considerations on data rate variants. In this simulation, it is analyzed that for internal users, 'External Load' has become a reason of decrease in throughput, an important QoS metric.

## 5. Conclusions

In this paper, we presented simulation results and performance analysis of Quality of Service (QoS) based on two major queuing disciplines i.e. Priority Queuing (PQ) and Weighted Fair Queuing (WFQ). The analysis was done in terms of delay and its variants for four different services; Voice over IP, Video Conferencing, Database and FTP based on packet identification under Differentiated Services Code Point (DSCP). The simulation results show that WFQ is a better discipline than PQ as lesser queuing delay and jitter were observed in WFQ for low priority services (FTP and Database). These metrics were found having same values in both PQ and WFQ for high priority services (Voice and Video). Therefore, low priority services also get their weighted share of bandwidth in the network when QoS is applied with WFQ discipline in the presence of high priority services like interactive voice and live streaming video. Voice and video traffic were further examined through simulations and discussion on 'Traffic Drop', 'Buffer Usage' and 'Packet Delay Variation' metrics by varying router's buffer size. In this way, we have been able to perform a deeper QoS analysis in the network for a variable parameter of router's interface. It was found when buffer size was tuned to 10 kilobytes, video packets ceased to drop and packet delay variations reduced significantly for both voice and video.

The same network was extended to include Internet based communication and VPN was configured to allow the access of 'Data Server' to the external user for Database service. It was observed when the server was accessed internally as well as by the external user, average data rate of Database traffic received by internal network users (in bytes/sec) decreased due to external load. Therefore, we are able to conclude that packet markings for Quality of Service are not enough to guarantee throughput of a service to internal network users when the traffic source is also shared by external load.

**Muhammad Aamir** is MS-IT (Computing) Student at Shaheed Zulfikar Ali Bhutto Institute of Science & Technology (SZABIST), Karachi, Pakistan. He has done his Graduation (Bachelors of Engineering in Industrial Electronics) from Institute of Industrial Electronics Engineering (IIEE) under affiliation of NED University of Engineering & Technology, Karachi, Pakistan. His topics of interest are Computer Networks, Communication Systems, Automation and Human Machine Interfacing. He is currently a member of IEEE and Computer Society of IEEE. He is also a reviewer of IEEE conference papers.

**Syed Mustafa Ali Zaidi** is a full time research scholar at the Faculty of Computing, Shaheed Zulfikar Ali Bhutto Institute of Science & Technology (SZABIST), Karachi, Pakistan. He received a Master of Applied Physics degree from Karachi University in 1994, Master of Power Engineering degree from NED University Karachi in 1997 and Master of Computer Science degree from SZABIST in 2010. He has received merit scholarship during the Master degrees and currently receiving scholarship for Ph.D. from Higher education commission Pakistan. His research interests include Computer Networks, Communication Systems and Optimization of multivariable quality characteristics using statistical & machine learning tools. He is also a member of IEEE.

**Dr. Husnain Mansoor Ali** is an Assistant Professor / MS-Coordinator at Department of Computing, Shaheed Zulfikar Ali Bhutto Institute of Science & Technology (SZABIST), Karachi, Pakistan. He received his B.E. Computer and Information Systems in 2004 from NED University Karachi, M.S. Networks and Telecom in 2006 and Ph.D. in 2010 from University of Paris SUD-XI. His research interest is in the area of Ad-Hoc Networks. He also has extensive experience of teaching and research.